\documentstyle[preprint,aps]{revtex}
\begin{document}

\draft
\preprint{$
\begin{array}{l}
\mbox{LBL-37426}\\ [-3mm]
\mbox{UCB-PTH-95/20}
\end{array}$}

\title{Reinterpretation of Thermal Dilepton
Emission \\ Rate
by Spectral Functions\thanks{
\baselineskip=12pt
This work was supported by the Director, Office of Energy
Research, Office of High Energy and Nuclear Physics, Division of High
Energy Physics of the U.S. Department of Energy under Contract
DE-AC03-76SF00098.}
}
\author{Zheng Huang\footnote{Zhuang@lbl.gov} }

\address{{
Theoretical Physics Group,
Lawrence Berkeley Laboratory\\
\baselineskip=12pt       University of California,
    Berkeley, California 94720, USA}}

\date{June 20, 1995}
\maketitle
\begin{abstract}
We reinterpret the dilepton emission rate from a hadronic
gas expected to be produced
in heavy ion collisions in terms of the spectral functions
available from the $e^+e^-$ annihilation and the $\tau$ lepton
decays experiments. We take into account all possible
hadronic state especially the multi-pion
contributions to the dilepton
emission and the parity mixing phenomenon due to the soft
final-state corrections. A new compilation of the experimental
data for the spectral functions is presented.
\end{abstract}
\newpage
\narrowtext

The lepton pair emissions are the ideal probes of the dense system of strongly
interacting particles expected to be produced in ultrarelativistic heavy
nucleus-nucleus collisions \cite{rev}.
Once produced, they will decouple from the strongly
interacting system without further interactions and thus carry the information
on the dynamical properties of the hot nuclear system. The theoretical
predictions on the emission rate, both from a quark-gluon plasma at high
temperature ($T$) and from a hot hadron gas at relatively lower $T$, have been
extensively discussed in literatures
\cite{mcle,ruus,cley,gale,slk}.
There already exist some experimental data from various CERN SPS experiments
(for a recent review, see \cite{tser95}), and more thorough investigations will
be made at upcoming RHIC and LHC experiments. The quantitative
interpretation of
the data is beginning to be possible \cite{sriv,xion}

The main theoretical uncertainties in calculating the dilepton emissions rate
come from the fact
(apart from the crucial assumption that the system thermalizes in short
time and
stays in  local equilibrium)  that the system is strongly
interacting and the perturbation
theory fails. In a hot hadronic gas, though the
effective degrees of freedom are mainly pions, pions interact strongly among
themselves in many resonance channels,
and effectively the resonances have to be
taken into account. One obvious example is the pion electromagnetic
form factor which exhibits the $\rho$-meson dominance. Near the resonances the
naive pole approximations often violate the unitarity of the partial waves.
There may also be some potential double counting problems associated with the
emission rate at the location of the resonance peaks, as to whether or not
the real resonance dilepton decays should be counted separately. Also, the
multi-pion state emissions, e.g.\ $3\pi\rightarrow \ell^+\ell^-$ and
$4\pi\rightarrow \ell^+\ell^-$ may not be so suppressed as they can proceed
through the resonance poles,
e.g.\ $3\pi\rightarrow \omega ^*,\phi ^*\rightarrow\ell^+\ell^-$
and   $4\pi\rightarrow \pi a_1\rightarrow\ell^+\ell^-$. Taking all these
reactions into account by the kinetic theory is a difficult task though not
impossible. One may always wonder the validity of a calculation and if the
omission of certain reactions oversimplifies the life.
More importantly, process
involving hadrons in the final state such as
$n\pi\rightarrow \ell^+\ell^- m\pi$ may have
been entirely overlooked. Even in
the phase of quark-gluon plasma where quarks and gluons are supposed to be
weakly interacting, the high order corrections in $\alpha_s$ may not be so
small.

The aim of this paper is to use and to develop the expression that allows the
non-perturbative treatment of the strong interactions and avoids the detailed
counting of the relevant reactions. Such an expression is known to exist
\cite{mcle,ruus,weld} in the form of a low temperature expansion when $T^2$
is smaller compared to the dilepton invariant mass squared $M^2$. We shall
reinterpret the dilepton emission rate in terms of the spectral functions
of some hadronic currents that are known experimentally in the low energy
$e^+e^-$ annihilation process and $\tau$ lepton decays.
While we shall comment
on its application to the quark-gluon plasma, our approach is most relevant
for the case of a hadronic gas. The recent development in calculating
the correlation functions of vector currents in a low temperature pion gas
using the PCAC and current algebra \cite{dei}
enables us to obtain the important
corrections to the rate with soft pions emitted together with
the dilepton in the final state. Our parameterization
clearly exhibits the mixing
phenomenon of vector and axial vector currents at finite temperature and its
relation to the dilepton production. We present a new compilation of the
experimental data for the spectral functions in the energy range from
360 MeV to 2150 MeV, relevant to the dilepton yield from a hadron gas.

In the lowest order in the electromagnetic interactions,
the emission rate can
be expressed non-perturbatively in terms of the correlation functions
of the hadronic electromagnetic current. When the lepton
mass is ignored, the rate reads \cite{mcle,ruus,weld}
\begin{eqnarray}
\frac{dR}{d^4q}=\frac{4\alpha^2}{3(2\pi)^3}\frac{1}{q^4}(q^\mu
q^\nu -q^2g^{\mu\nu})W_{\mu\nu}(q) \label{1}
\end{eqnarray}
where
\begin{equation}
W_{\mu\nu}(q)=\int d^4x e^{-iqx}\langle\langle J_{\mu}^{\rm em}(x)
J_\nu^{\dagger {\rm em}}(0)\rangle\rangle \; .
\end{equation}
$q$ is the time-like 4-momentum of the lepton pair, $J_{\mu}^{\rm em}$ is
the electromagnetic current for hadrons and $\langle\langle \cdots
\rangle\rangle$ stands for the thermal average at a give temperature $T$.
Note that $W_{\mu\nu}(q)$ is the thermal structure function, and there is
no time ordering in the product of $J_{\mu}^{\rm em}(x)$ and
$J_{\nu}^{\dagger {\rm em}}(0)$, which is crucial in relating $W_{\mu\nu}$
directly to the spectral functions. (\ref{1}) is shown to agree with the
relativistic kinetic theory reaction by reaction if one inserts the complete
set $\sum_F |F\rangle\langle F|=1$ for each component $|I\rangle$ of the
thermal density matrix
\begin{equation}
W_{\mu\nu}(q)=\sum_F\sum_I\int d^xe^{-iqx}\langle I|J_{\mu}^{\rm em}(x)
|F\rangle\langle F| J_\nu^{\dagger {\rm em}}(0)|I\rangle \frac{e^{-\beta
E_I}}{\cal Z}\; , \label{3}
\end{equation}
where ${\cal Z}$ is the canonical partition function of the system.
(\ref{3}) describes the all possible transitions
$I\rightarrow F \ell^+\ell^-$. Summing up the disconnected parts of
the matrix
elements, one recovers the appropriate  Fermi or Bose distribution
functions $f(E_i)=(e^{\beta E_i}\pm 1)^{-1}$ for each initial-state
particle and the suppression or enhancement factor $(1\mp f(E_i))$ for each
final-state particle. $W_{\mu\nu}(q)$ has no non-trivial zero temperature
limit. To get the leading temperature dependence, one alternatively sums up
all initial states $|I\rangle$'s using the completeness $\sum_I |I\rangle
\langle I|$ and the energy conservation $E_I=E_F+q^0$ \cite{weld}
\begin{eqnarray}
W_{\mu\nu}(q) & = & e^{-\beta q^0}\sum_F \int d^4xe^{iqx}
\langle F|J_\nu^{\dagger {\rm em}}(x) J_\mu^{{\rm em}}(0)|F\rangle
\frac{e^{-\beta E_F}}{{\cal Z}}\nonumber \\
 & =&  e^{-\beta q^0} \int d^4xe^{iqx}
\langle\langle J_\nu^{\dagger {\rm em}}(x) J_\mu^{{\rm em}}(0)\rangle\rangle
\equiv e^{-\beta q^0}\rho_{\mu\nu}(q)\; . \label{4}
\end{eqnarray}
$\rho_{\mu\nu}(q)$ now has a non-trivial limit as $T\rightarrow 0$.
Though $|F\rangle$'s and $|I\rangle$'s
seem to switch their positions in (\ref{3}) and (\ref{4}),
the same kinetic theory result would be obtained if one uses the identity
$(1\mp f(E))e^{-\beta E}=f(E)$.

We would like to compute $\rho_{\mu\nu}(q)$ for a low temperature hadronic
gas, not reaction by reaction as one does
in the kinetic theory, but to obtain a
low $T$ expansion, in which each term is the summation of all possible
reactions. The leading term is independent of temperature, which is
obtained by taking the limit $T\rightarrow
0$ where only the vacuum state in $|F\rangle$'s contributes to the
connected part, namely
\begin{equation}
\rho_{\mu\nu}(q)\stackrel{T\rightarrow 0}{\longrightarrow}
\rho_{\mu\nu}^{\rm em}(q)\equiv \int d^4x e^{iqx} \langle 0|
J_\nu^{\dagger {\rm em}}(x) J_\mu^{{\rm em}}(0)|0\rangle\; .
\label{5}
\end{equation}
When all possible initial states $|I\rangle$'s are saturated between
the two currents, (\ref{5}) corresponds to the ``hard'' particle
contributions illustrated  in \cite{mcle} where ``hard'' means ``not
necessarily soft''. As we are interested in the kinematical region
$T^2\ll q^2$, any non-vacuum states $|F\rangle$'s are necessarily
soft since they are kinematically limited to $E\leq T$ by the presence
of a thermal distribution function. The low but finite temperature
corrections to the leading term $\rho_{\mu\nu}^{\rm em}(q)$ arise
from the soft final-state particle emissions and the soft initial-state
particles, both are limited to $E\leq T$. In a hadronic gas, the soft
modes are the pions. One may use the soft pion theorem based on the PCAC
to calculate the soft pion corrections. Unlike the case in the
quark-gluon plasma where there is a contribution of order $T^2$, the
insertion of the soft pions (the initial states) in the intermediate
states of the correlator does not pick up a contribution of order $T^2$.
In fact, the extra powers of soft thermal pion momentum leads to a
suppression factor of order $T^4/M^4$ in the chiral limit, as shown by
Leutwyler and Smilga in \cite{leut} and Eletsky in \cite{dei}. This is
the consequence of the low temperature theorem \cite{leut} (the finite
temperature version of the Adler theorem) that the chiral symmetry
protects all of the masses from picking up a contribution of order
$T^2$. We shall ignore the effects of the initial-state soft corrections
with the understanding that they start at $O(T^4/M^4)$.

On the other hand, the corrections from the soft final-state thermal pions,
as shown by Dey, Eletsky and Ioffe \cite{dei}, leads to the mixing
of vector and axial vector correlators. The essence is to approximate
the matrix elements containing soft pions with their threshold value
at zero pion momentum, which is completely determined by the
current algebra or the symmetry of the system \cite{shur}. For
example,
\begin{eqnarray}
\langle n\pi^+|J_\nu^{\dagger {\rm em}}(x) J_\mu^{{\rm em}}(0)|n\pi^+
\rangle & = & -\frac{1}{F_\pi^2}\langle (n-1)\pi^+|J_\nu^{\dagger
{\rm em}}(x)
V^3_\mu (0)-A^-_\nu (x)A^+_\mu (0)\nonumber\\
& &-A^+_\nu (x)A^-_\mu (0)
+V^3_\nu J_\mu^{{\rm em}}(0)|(n-1)\pi^+\rangle \label{6} \\
& =& \frac{2}{F_\pi^2}(-1)^n3^{n-1}\langle 0|V^3_\nu (x)V^3_\mu (0)
-A^3_\nu (x) A^3_\mu (0)|0\rangle\; , \nonumber
\end{eqnarray}
where $V^3_\mu$ and $A^3_\mu$ are the third (neutral) component of the
vector- and axial-vector- isovector currents, $F_\pi=93$ MeV.
By doing so, we may compute all reactions with soft pions in the
final state $I\rightarrow \ell^+\ell^-n\pi^\pm$. It is worth
pointing out that although $J_\mu^{{\rm em}}$ contains all flavors ($u,d,
s,c,b,t$), the soft pion emission selects only two flavors ($u,d$) since
$V^3_\mu$ and $A^3_\mu$ contains no strange or higher flavor contents.
Also, when only $u,d$ flavors are effective at lower energy range,
$J_\mu^{{\rm em}}\equiv V^3_\mu +B_\mu/2$ where $B_\mu$ is the baryonic
current which contains both isovector ($I=1$) and isosinglet ($I=0$)
parts, while $V^3_\mu$ and $A^3_\mu$ are purely isovector. When $n\geq 2$,
(\ref{6}) combining with the integration of the Bose distribution functions
gives the corrections of order $(T^2/6F_\pi^2)^2$ or higher. However, the
multi-pions in the final state can interact with themselves and cause
further phase shifts. Inclusion of the $\pi\pi$ rescattering effects gives
a correct $T^4$ term \cite{dei}. We neglect the non-zero pion momentum that
causes a correction suppressed by $M^2$ and
summarize the corrections to $T^4$ in the
spectral tensor
\begin{equation}
\rho_{\mu\nu}(q)=\rho_{\mu\nu}^{\rm em}(q)-(\epsilon -\frac{\epsilon^2}{2})
(\rho_{\mu\nu}^{\rm V}(q)-\rho_{\mu\nu}^{\rm A}(q))\; , \label{7}
\end{equation}
where $\epsilon =T^2/6F_\pi^2$. $\rho_{\mu\nu}^{\rm V}(q)$ and
$\rho_{\mu\nu}^{\rm A}(q)$ are the correlators defined in (\ref{5}) with
$J_\mu^{\rm em}$ replaced with $V^3_\mu$ and $A^3_\mu$ respectively.
Clearly, the dilepton production from a thermal source comes not only
from the vector channel but also from the axial vector channel since
they can interchange their parity identities through the interaction
with a thermal background.
The soft corrections tend to suppress the contributions from the
 vector channel
and to increase the rate by adding to it the axial channel contributions.

Since $J_\mu^{\rm em}$ and  $V^3_\mu$ are conserved currents (neglecting
the isospin breaking), $\rho_{\mu\nu}^{\rm em}(q)$
and $\rho_{\mu\nu}^{\rm V}(q)$ have only transverse (spin one) part.
 $A^3_\mu$ receives both spin one and spin zero (pions) contributions.
However, the longitudinal part ($\propto q^\mu q^\nu$) does not contribute
when contracted with the lepton pair tensor. Define the spectral
functions $\rho^{(i)}(s)$ ($i={\rm em, V, A}$) \cite{shur}
\begin{equation}
\rho_{\mu\nu}^{(i)}(q)=2\pi (q_\mu q_\nu -q^2g_{\mu\nu})\theta (q^0)
 \rho^{(i)}(s) \; ,\label{8}
\end{equation}
where $s=q^2$, one obtains the dilepton emission rate in terms of three
spectral functions
\begin{equation}
\frac{dR}{dq^4}=\frac{4\alpha^2}{(2\pi)^2}e^{-\beta q^0}
\left \{ \rho^{\rm em}(s)-(\epsilon -\frac{\epsilon^2}{2})[
\rho^{\rm V}(s)-\rho^{\rm A}(s)]\right \}\; , \label{9}
\end{equation}
or if the transverse mass and the rapidity are integrated over
\begin{equation}
\frac{dR}{dM^2}=\frac{4\alpha^2}{2\pi}MTK_1(M/T)
\left \{ \rho^{\rm em}(M)-(\epsilon -\frac{\epsilon^2}{2})[
\rho^{\rm V}(M)-\rho^{\rm A}(M)]\right \}\; . \label{9b}
\end{equation}
As we have noted earlier, because the spectral tensor
 does not involve the time
ordering of the currents, it is directly represented by the spectral functions
on the physical axis $q^2>0$. In the case of time ordering correlator,
the spectral function is (twice) the imaginary
part of the correlator. To get the real part, one has to perform the
dispersion
integral of the spectral function. Clearly, the usual pole-plus-continuum
approximation widely used in the Weinberg's sum rules \cite{shur}
would not do since
it is precisely the energy dependence of the spectral densities that determine
the shape of the dilepton spectrum.

Other than a theoretical estimate, a more sound procedure is to use
experimental date which have been accumulated over years and measured
accurately enough to truly represent these spectral functions. Obviously,
$\rho^{\rm em}(s)$ can be extracted from the total hadronic cross section
(the $R$ value) of $e^+e^-$ annihilations (note that
$\sigma ({\rm had}\rightarrow e^+e^- )=4\sigma (e^+e^-\rightarrow {\rm had})$
because of the helicity summation)
\begin{equation}
\rho^{\rm em}(s)=
\frac{s\sigma (e^+e^-\rightarrow {\rm had})}{16\pi^3\alpha^2}
=\frac{R(s)}{12\pi^2}\; .\label{10}
\end{equation}
To get $\rho^{\rm V}(s)$, we have to first restrict the open flavors to
$u,d$. This can be done effectively by selecting the events
in which only pions
are produced. We also need to eliminate the $I=0$ events in $e^+e^-$
annihilations. Since the $I=0$ states will have the $G$ parity
$G=C(-1)^I=-1$,
which can only decay into an odd number of pions (a pion has $G=-1$). So by
selecting the events consisting only of even number of pions, we have for
$\rho^{\rm V}$ \cite{pesk,dono}
\begin{equation}
\rho^{\rm V}(s)=
\frac{s}{16\pi^3\alpha^2}\sum_{n=1}\sigma (e^+e^-\rightarrow 2n\pi )
\; .\label{11}
\end{equation}
$ \rho^{\rm A}$ can be extracted from the differential probabilities
of the $\tau$ lepton decays into odd number of pions
$\tau \rightarrow \nu_\tau (2n+1)\pi$ \cite{tsai}
\begin{equation}
\rho^{\rm A}(s)=\frac{8\pi m_\tau^3}{G_F^2\cos ^2\theta_c (m_\tau^2+2s)
(m_\tau^2-s)^2}\sum_{n=1}\frac{d\Gamma (\tau \rightarrow \nu_\tau
(2n+1)\pi )}{ds}\; ,\label{12}
\end{equation}
where $s$ is understood to be the invariant mass squared of $(2n+1)\pi$'s,
$G_F$ is the Fermi constant and $\theta_c$ is the Cabibbo angle.

We present a new compilation of the data for those spectral functions in
the energy range fro 360 MeV and 2150 MeV where the data are rich. The total
hadronic cross section of $e^+e^-$ annihilation above $\sqrt{s}=1420$ MeV is
measured by Adone $\gamma\gamma 2$ and MEA \cite{adone79},
below which the ORSAY data exist \cite{wiik} but are fairly sparse,
especially
at $\rho ,\omega , \phi$ peaks. We fill in more data from
OLYA and CMD $e^+e^-\rightarrow 2\pi$ measurement \cite{olya},
Orsay DM1 $e^+e^-\rightarrow 3\pi$ measurement \cite{dm1}, and VEPP-2M
$e^+e^-\rightarrow K^+K^-, K_SK_L$ \cite{vepp} in the resonance region.
To extract $\rho^{\rm V}$, we consider, in addition to
$e^+e^-\rightarrow 2\pi$, the multi-pion reactions
$e^+e^-\rightarrow 2\pi^+2\pi^-$, $\pi^+\pi^-2\pi^0$,
$2\pi^+2\pi^-2\pi^0$, $3\pi^+3\pi^-$ which are available from Adone
\cite{adone2}. For $\rho^{\rm A}$, we assume the equality of
$\Gamma (\tau \rightarrow \nu_\tau \pi^+2\pi^-)$ and
$\Gamma (\tau \rightarrow \nu_\tau \pi^-2\pi^0)$ which is expected if the
$3\pi$ state is dominated by the $a_1(1230)$ resonance, and use
the recent ARGUS data \cite{argus}. Clearly, $\rho^{\rm A}$ thus
obtained is limited to $\sqrt{s}<m_\tau\sim 1.7$ GeV. However, the rapid fall
off of $\rho^{\rm A}$ beyond $m_{a_1}$ indicates a rather negligible
contribution at a larger $\sqrt{s}$.

We plot our compilation of the numerical values of
the three spectral functions in Fig.\ 1 for the convenience of further
phenomenological studies.  Data with poor statistics are removed by hand.
The dilepton emission rate at a typical temperature $T=160$ MeV is plotted
in Fig.\ 2 for the invariant mass ranging from 360 MeV to 2150 MeV.
The dotted line is the leading
contribution without the soft final-state emissions. An enhancement located
around $m_{a_1}\simeq 1230$ MeV is visible due to $\rho^{\rm A}$.

We have
reinterpreted the dilepton emission rate in terms of spectral functions
available directly from the data. The leading contribution is related to the
total inclusive hadronic cross section of $e^+e^-$ annihilation, which sums
up all possible combinations of mesons and resonances such as
$P+P$, $V+V$, $V+P\rightarrow \ell^+\ell^-$ recently considered by
Gale and Lichard \cite{gale}, and three- and four- body combinations, and many
more. It is free of theoretical uncertainties in calculating these
non-perturbative processes and avoids the possible double counting of the
light mesons and their resonances. As emphasized by McLerran and Toimela
\cite{mcle}, the leading term by no means contains only the hard particle
contributions. It is merely not necessarily soft which includes the soft
particle contributions as well. The parity mixing phenomenon, also considered
in \cite{slk,xion,hsw}
is explicit in our reinterpretation by the presence of the axial spectral
density $\rho^{\rm A}$, which arises from the soft final-state emission
of pions. Our formalism is applicable to a hadronic gas as long as
$T^2\ll M^2$, $T^2\ll 6F_\pi^2$ and $T<T_c$. It may be possible to extend the
approach to the case of quark-gluon plasma for the high invariant mass region
$M>2$ GeV, where $\sigma (e^+e^-\rightarrow {\rm had})\simeq
\sigma (e^+e^-\rightarrow q\bar q)$ and $\alpha_s$ corrections to all orders
would be automatically included . When $M<2$ GeV,
the parton degree of freedom
are not apparent in $\sigma (e^+e^-\rightarrow {\rm had})$. The existence
of the parton degrees freedom at $T>T_c$ even for low invariant mass parton
pair is a fundamental assumption.
To calculate the actual dilepton multiplicity in order to comapare
it with the experimental data, one has to integrate the rate over the
space-time  history of the thermal source, which involes more assumptions
and approximations, and will be reported elsewhere.

I would like to thank Y.\ Kluger, A.\ Kovner and X.-N.\ Wang
 for useful discussions.
This work was supported in part by the Director, Office of
Energy Research, Office of High Energy and Nuclear Physics, Division of
High Energy Physics of the U.S. Department of Energy under Contract
DE-AC03-76SF00098
and  by the Natural Sciences and
Engineering Research Council of Canada.

\newpage
\centerline{\bf Figure Captions}
\vskip 15pt
\begin{description}
\item[Fig. 1] Spectral functions $\rho^{\rm em}(s)$, $\rho^{\rm V}(s)$,
and $\rho^{\rm A}(s)$ from $e^+e^-$ annihilation and $\tau$ decays.
Note that $\rho^{\rm em}(s)$ extends beyond the vertical scale at the
$\omega$ and $\phi$
resonance peaks.
\item[Fig. 2] Dilepton emission rate from a $T=160$ MeV thermal source.
Dotted line (error bars not shown)
represents the leading temperature contribution without the
soft pion emissions. The error bars are calculated assuming the statistical
independence of different experiment sets.
\end{description}
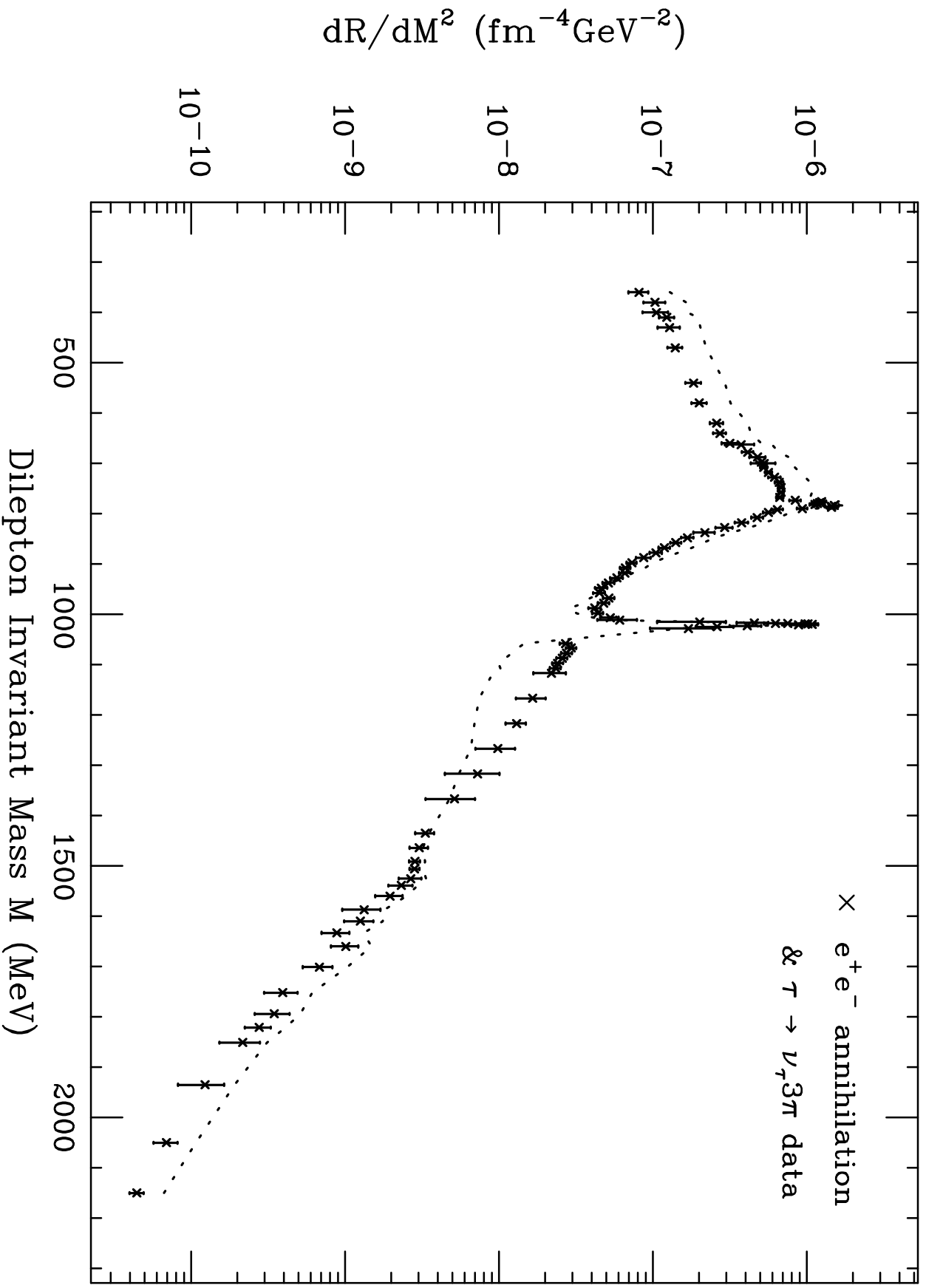
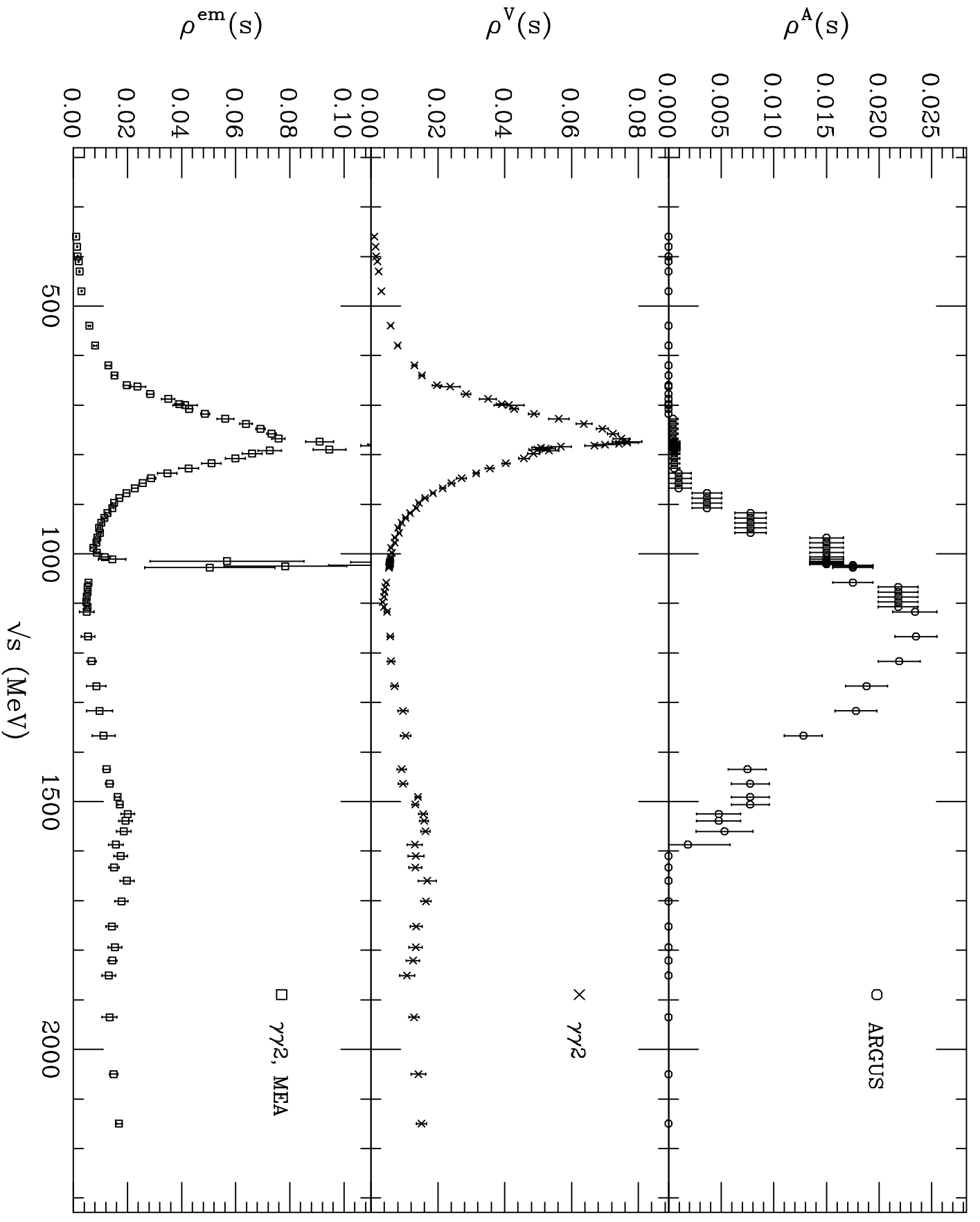
\end{document}